\documentclass[showpacs,aps,amsmath,amssymb]{revtex4}

\usepackage{graphicx}
\usepackage{ulem}
\usepackage{bm}

\begin{document}

%\wideabs{

\title{Rapid rotation of a Bose-Einstein condensate in a harmonic plus quartic
trap}

\author{Alexander L. Fetter$^{1}$, B. Jackson$^{2}$, and
   S. Stringari$^{2}$}

   \affiliation{$^1$Geballe Laboratory for Advanced Materials and Departments of
   Physics and Applied Physics,\\ Stanford University, California
94305-4045, USA\\
$^2$Dipartimento di Fisica, Universit\`a di Trento and BEC-INFM,
   I-38050 Povo, Italy.}

\date{\today}

\begin{abstract}

A two-dimensional  rapidly rotating Bose-Einstein condensate in an
anharmonic trap with quadratic and quartic radial confinement is
studied   analytically with the Thomas-Fermi approximation and
numerically with the full time-independent Gross-Pitaevskii equation.
The quartic trap potential allows the rotation speed $\Omega$ to
exceed the radial harmonic frequency $\omega_\perp$.  In the regime
$\Omega \gtrsim  \omega_\perp$, the condensate contains a dense
vortex array (approximated as solid-body rotation for the analytical
studies).  At a critical angular velocity $\Omega_h$, a central hole
appears in the  condensate.  Numerical studies confirm the predicted
value of $\Omega_h$, even for interaction parameters that are not in
the Thomas-Fermi limit.  The behavior is also investigated at 
larger angular
velocities, where the system is expected to undergo a transition to a giant
vortex (with pure irrotational flow).

\end{abstract}

\maketitle

\section{Introduction}

The development of experimental techniques
to create a single vortex in a dilute trapped Bose-Einstein
condensate (BEC)~\cite{Matt99,Madi00} rapidly led to larger arrays
containing up to several hundred
vortices~\cite{Abo01,Rama01,Halj01,Enge02}.  Typically, these
condensates  rotate rapidly, with angular velocities $\Omega$ that
approach the radial trap oscillator frequency $\omega_\perp$.    The
resulting centrifugal effect significantly weakens the radial
confinement, so that the condensate expands radially and shrinks
axially~\cite{Rama01,Halj01,Fett01}.

In a pure harmonic radial trap,
the limit $\Omega\to \omega_\perp$ is singular, because the
Thomas-Fermi (TF) radius diverges and the central density decreases
toward zero.  Consequently, the angular momentum also diverges,
and it becomes increasingly difficult to spin up the condensate   as it
approaches this limit.  In addition, the single-particle
Hamiltonian reduces to that of a
charged particle in a uniform magnetic field~\cite{Ho01}, and the
ground state can be constructed from the lowest-Landau-level wave
functions.  Recent experiments have explored the crossover region
between these two regimes~\cite{Schw04,Codd04}.

One procedure that
eliminates this singularity is to add a stronger radial  potential
that confines the condensate even for $\Omega \ge \omega_\perp$. A
quartic potential provides a particularly simple
form~\cite{Fett01,Lund02} that has recently been realized
experimentally with a blue-detuned laser
directed along the axial direction~\cite{Bret04,Stoc04}.

For this combined potential, the system is expected to have
no vortices at sufficiently slow rotation speeds $\Omega$.  With
increasing $\Omega$, there is  a sequence of states with  an increasing
number of vortices that eventually form a
relatively large vortex lattice.   This behavior is not qualitatively different
from that in
a pure harmonic trap.  The new  feature of the quartic confining
potential is  that the condensate continues to expand radially for
$\Omega >\omega_\perp$, with a central hole appearing at a critical
angular velocity $\Omega_h$. For still larger values of $\Omega$, the
condensate has an annular form with a mean radius that continues to
expand with increasing $\Omega$.
Ultimately, a giant vortex is expected to appear when the
singly quantized vortices disappear from the Thomas-Fermi condensate,
leaving an axisymmetric state with pure irrotational flow in the
annulus.  One reason for the interest in the additional  quartic
potential is that the central hole and giant vortex do not occur for
a pure harmonic trap potential.

Previous theoretical work on this system has been mainly
numerical~\cite{Kasa02,Kavo03}  or for
weak interactions and small anharmonicities~\cite{Jack04}, and
  here we report an
analytical
study of the TF regime,  approximating the actual superfluid velocity
of the dense vortex array by solid-body rotation $\bm {v}_{\rm sb} =
\bm \Omega\times \bm r$.  In addition, we perform  a full numerical
study of the two-dimensional Gross-Pitaevskii equation  for
different interaction strengths.  Comparison of the numerical and
analytical  studies confirms the qualitative features of the  TF
analysis.  Section II summarizes the basic TF procedure.  This
approach predicts the critical angular velocity $\Omega_h$ for  the
initial appearance of a central hole, surrounded by a dense vortex
lattice (Sec.~III).  The corresponding numerical study is described
in Sec.~IV, confirming the TF result for $\Omega_h$.  For larger
values of  $\Omega > \Omega_h$,  the TF approximation predicts that
the annular condensate expands radially with constant area.  Thus the
width of the annulus decreases, and eventually the vortices are
expected to retreat into the central hole (Sec.~V), leaving a pure
irrotational state with macroscopic circulation (often called a
``giant vortex'')~\cite{Kasa02,Kavo03,Fisc03}. Since solid-body
rotation always minimizes the energy in the rotating frame,   this
transition to a giant vortex depends essentially on the discrete
character of the quantized superfluid vortices.  Comparison with  our
numerical studies shows that the theoretical analyses are less
successful in modeling this
transition~\cite{Kavo03}.

\section{Summary  of procedure}

The  problem of interest is  the equilibrium state of a rapidly
rotating trapped
Bose-Einstein condensate in a two-dimensional confining potential
that has both quadratic
(harmonic) and quartic components.  In terms of the usual dimensional
quantities, the trap potential has  the  form
\begin{equation}
V_{\rm tr}(r) = \frac{1}{2}M\omega_\perp^2\left(r^2 +
\lambda\frac{r^4}{d_\perp^2}\right) = \frac{1}{2}\,\hbar
\omega_\perp\left(\frac{r^2}{d_\perp^2} +
\lambda\frac{r^4}{d_\perp^4}\right),
\end{equation}
where $d_\perp=\sqrt{\hbar/M\omega_\perp}$ is the harmonic-oscillator
length, $\bm  r$ is the two-dimensional radial coordinate and $\lambda$ is
a dimensionless  constant that characterizes the admixture of the
quartic component.  For simplicity, it is convenient to
treat  an effectively  two-dimensional system, uniform in the $z$
direction over a length
$Z$. Kavoulakis and Baym~\cite{Kavo03} considered
the same two-dimensional system, especially in the limit of rapid rotation
where the condensate develops a central hole and becomes  annular.

In a frame  rotating with angular velocity $\Omega$, the energy is given
by the functional
\begin{equation}\label{eprime}
E' = \int dV\,\left[\frac{\hbar^2}{2M}\,|{\bm\nabla}\Psi|^2 +
V_{\rm tr}(r)\,|\Psi|^2 +\frac{2\pi \hbar^2 a}{M}\,|\Psi|^4-\Omega\,L_z\right],
\end{equation}
where $\Psi$ is the condensate wave function, the integral is over
the three-dimensional volume, $a>0$ is the $s$-wave scattering length
and
$L_z=\Psi^* [\hat{\bm z}\cdot (\bm {r}\times{\bm  p})]\Psi$
is the
$z$ component of  angular momentum.  Once $E'$ is evaluated for fixed
particle number $N = \int dV |\Psi|^2$ and fixed
$\Omega$, the  chemical
potential and angular momentum follow from the thermodynamic relations
\begin{equation}\label{thermo}
\left(\frac{\partial E'}{\partial N}\right)_\Omega =
\mu, \qquad \left(\frac{\partial E'}{\partial \Omega}\right)_N =
-L_z.
\end{equation}
Equivalently, the free energy $F = E' -\mu N$
incorporates the constraint of fixed $N$ with the chemical potential
$\mu$ as a Lagrange
multiplier.

For the specific two-dimensional case studied here, the condensate wave
function can be chosen as $\Psi = \sqrt{N/Z}\,\psi({\bm r})$, and the
normalization condition becomes
\begin{equation}\label{norm}
\int d^2 r\,|\psi|^2 =1.
\end{equation}
It is convenient to use dimensionless harmonic-oscillator units, with
$\omega_\perp$ setting
the scale for frequency and energy, and $d_\perp$ setting the
scale for length. In addition, we use a dimensionless coupling
parameter $g = 4\pi N a/Z$, so that Eq.~(\ref{eprime}) has the
dimensionless form
\begin{equation}\label{edim}
E' = \int d^2 r \,\left[\frac{1}{2}|{\bm \nabla}\psi|^2 +
\frac{1}{2} \left(r^2 + \lambda r^4\right)\,|\psi|^2  +\frac{1}{2}
g\,|\psi|^4 - \Omega L_z\right],
\end{equation}
while Eq.~(\ref{norm}) is unchanged.
Using the replacement $\psi=\sqrt{n}\,{\rm e}^{iS}$, we 
can rewrite the
first term in Eq.~(\ref{edim}) as $|{\bm \nabla}\psi|^2=|{\bm \nabla}
\sqrt{n}|^2 +|{\bm \nabla} S|^2 |\psi|^2$. The Thomas-Fermi approximation
will be used here, which corresponds to neglecting the curvature of the
density $n$. This contribution  becomes significant only
near to the edge of
the condensate and at vortex cores, and we will discuss the validity
of this approximation later. The usual
expression for the dimensionless superfluid density
${\bm v}={\bm \nabla} S$ as a gradient of the phase gives
\begin{equation}\label{etf}
E' = \int d^2 r \,\left[
\frac{1}{2} \left(v^2 + r^2 + \lambda r^4\right)\,|\psi|^2   -
\bm\Omega\,\cdot \bm r\times
\bm v\,|\psi|^2 +\frac{1}{2}g\,|\psi|^4 \right].
\end{equation}
Variation of the free energy with respect to $|\psi|^2$ yields the
familiar TF density
\begin{equation}\label{tfdens}
g\, |\psi|^2  = \mu +\bm\Omega\,\times \bm r\cdot
\bm v -\frac{1}{2}
\left(v^2 + r^2+\lambda\,r^4 \right),
\end{equation}
which allows a direct determination of $\mu$ from the normalization
condition (\ref{norm}) and $E'$ from (\ref{etf}); as a check,
$\mu(N,\Omega)$ must also
follow from the first thermodynamic relation in (\ref{thermo}).

The aim is to  study the system for
different annular condensates.

\begin{enumerate}

\item  The first is an annulus containing a
vortex lattice, for which  the superfluid flow is approximated   as
solid-body rotation $\bm v_{\rm sb} = \bm{\Omega\times r}$. This
approach has the principal advantage that all the physical properties
can be evaluated analytically, in contrast to the more elaborate
trial wave function and numerical approach used in
Ref.~\cite{Kavo03}.  The approximation of replacing the vortex
lattice by  uniform vorticity works well when the Wigner-Seitz
circular cell radius  $l = 1/\sqrt \Omega$ (in dimensional units, $l
= \sqrt{\hbar/M\Omega}$) is much smaller than all the dimensions of
the annulus.

\item As $\Omega$ increases, the annular condensate
becomes increasingly narrow, with the width $d$ proportional to $
1/\Omega$.  Eventually $d$  becomes comparable with the intervortex
spacing $\approx 2 l$.  Near this angular velocity,  elementary
considerations~\cite{Fett67} suggest that the vortex lattice would
disappear, indicating a transition to a giant vortex with
{\it irrotational} flow enclosing the central hole.  Note that  this
transition depends crucially on the nonzero quantum of circulation
$\kappa = h/M$, which determines  $l=\sqrt{\hbar/M\Omega}$.  A
quantitative  study of this transition first requires an analysis of
an annulus with pure irrotational flow, when the condensate wave
function has the form $\psi = |\psi|  e^{i \nu \phi}$, with $\nu$ a
large integer.  The irrotational velocity is  azimuthal with $v_{\rm
irr} = \nu/r$.

\item A detailed  theory of  the critical angular
velocity for the  transition to a giant vortex requires the inclusion
of the additional energy  associated with the local deviation of the
velocity from solid-body rotation in the vicinity of an individual
vortex core~\cite{Kavo03,Fisc03,Baym04}.    Previous numerical
studies of the Gross-Pitaevskii equation suggest that the relevant
comparison state in   a narrow annular condensate is  a
one-dimensional vortex array combined with macroscopic
circulation~\cite{Kasa02,Afta03}.  For simplicity, our  present
theoretical analysis treats only a single vortex at the midpoint of
the annular gap.  This  TF  model  yields a  value $\Omega_g$ for
this transition that is considerably larger than that predicted by
Kavoulakis and Baym~\cite{Kavo03}, but our numerical results indicate
that the actual transition occurs at still  larger
values.

\end{enumerate}

   \section{uniform vortex lattice
with central hole}\label{seclh}

   A single vortex located at the point ${\bm r}_0$ induces a
circulating flow velocity ${\hat {\bm z}} \times \left({\bm r - \bm
r}_0\right) /\left|{\bm r- \bm r}_0\right|^2$.  Thus, an array of
vortices located at the points $\{ {\bm r}_j\}$ induces a total flow

   \begin{equation}
\bm v(\bm r)  = \sum_j \frac{\hat {\bm z}\times \left(\bm r-\bm
r_j\right)}{\left| \bm r - \bm r_j\right|^2}.
\end{equation}
In the
limit of a dense vortex array with dimensionless areal density
$\Omega/\pi$ ($ = M\Omega/\pi\hbar$ in conventional units), the sum
can be approximated by an integral over the area of the annular
region $R_1 \le r \le R_2$, yielding
   \begin{equation}
\bm v_a(\bm r)  = \left(\Omega\, r  -\frac{\Omega
\,R_1^2}{r}\right)\hat{\bm \phi}
\end{equation}
for the  velocity
induced by the vortices in the annular region.
The first term is the
expected  solid-body rotation, but the second term
represents the hole
in the center of the annulus.  Thus it is necessary to add the
irrotational flow $\bm v_{\rm irr}(r) = (\Omega R_1^2/r)\hat{\bm
\phi} $ arising from the phantom vortices in the hole.  In the TF
approximation, these vortices  can be considered either to have the
same uniform density $\Omega/\pi $ or to combine into a single
macroscopic circulation with $\Omega R_1^2 $ quanta concentrated at
the origin, since the interior of the annulus lies outside the
physical region.  Thus the total superfluid flow  velocity in the
annular region becomes $\bm v = \bm v_a + \bm v_{\rm irr} =
\bm\Omega\times\bm r = \Omega r \hat{\bm\phi}$.

The solid-body flow velocity greatly simplifies the TF density in
Eq.~(\ref{tfdens}) to yield
\begin{equation}\label{dens1}
g |\psi|^2  = \mu  +\frac{1}{2}
\left[\left(\Omega^2-1\right) r^2-\lambda\,r^4 \right].
\end{equation}
For $\Omega < 1$, the density has a local maximum near
the center,  but it changes to a local minimum for $\Omega > 1$.  The
central density is proportional to $\mu$, which decreases
continuously with increasing $\Omega$~\cite{Fett01}.  For
sufficiently large rotation speeds,  the central density and $\mu $
  can  both vanish,  which corresponds to the point at
which the central hole first appears, as discussed in the next subsection.

\subsection{Onset of formation of a central hole}

The simple form of Eq.~(\ref{dens1}) allows a direct solution for
the  ``classical''  turning points where the TF density vanishes
\begin{equation}
R_i^2 = \frac{\Omega^2-1}{2\lambda} \pm \sqrt{\left(
\frac{\Omega^2-1}{2\lambda}\right)^2 +\frac{2\mu}{\lambda}},\qquad i =
1,2.
\end{equation}
Here the upper (plus) sign  denotes the outer squared radius $R_2^2$ for
any value of the chemical potential $\mu$. In contrast, the lower
(minus) sign yields a physical inner radius $R_1^2$ only if $\mu$ is
negative.

The central  hole in the uniform vortex lattice first appears when $\mu =
0$ (so that $R_1^2$ also vanishes).   For this value, the
normalization condition (\ref{norm}) gives the explicit
condition $12 g  = \lambda \pi R_2^6$.  This equation determines the
critical rotation frequency $\Omega_{h}$
  at which the central density
first vanishes (namely  the first appearance of a central
hole in the vortex lattice).  It can be written in the equivalent forms
\begin{equation}\label{omegalh}
12 g \lambda^2 = \pi \left(\Omega_{h}^2-1\right)^3,\qquad
\Omega_{h}^2 = 1 + 2\sqrt\lambda \left(\frac{3\sqrt\lambda \,g}{2\pi
}\right)^{1/3}.
\end{equation}
Note that $\Omega_{h}$ always exceeds 1 (namely
$\Omega_{h} >\omega_\perp$ in dimensional units).

\subsection{Properties of the vortex lattice with central hole}

If $\Omega>\Omega_{h}$, then the chemical potential  is
negative, and the squared TF  radii obey the simple relations
\begin{equation}\label{sd}
R_2^2 + R_1^2 = \frac{\Omega^2-1}{\lambda},\qquad R_2^2-R_1^2 =
      \sqrt{\left(
\frac{\Omega^2-1}{\lambda}\right)^2 -\frac{8\,|\mu|}{\lambda}}.
\end{equation}
The normalization condition (\ref{norm}) yields
$\lambda \pi \left(R_2^2-R_1^2\right)^3 = 12 g$, and comparison with
Eqs.~(\ref{omegalh}) and (\ref{sd}) immediately gives the (negative)
chemical potential
\begin{equation}\label{mulh}
\mu = \frac{\left(\Omega_{h}^2-1\right)^2 -
\left(\Omega^2-1\right)^2}{8\lambda}\qquad\hbox{for $\Omega >
\Omega_{h}$}.
\end{equation}
Equations~(\ref{sd}) show that the mean squared radius
grows with increasing angular velocity,
while, in contrast, the difference remains fixed
\begin{equation}\label{difflh}
R_2^2 -R_1^2 = \frac{\Omega_{h}^2-1}{\lambda} =
\frac{2}{\sqrt\lambda} \left(\frac{3\sqrt\lambda\, g}{2\pi}\right)^{1/3};
\end{equation}
for all $\Omega>\Omega_{h}$.  This relation shows that the area
$\pi\!\left( R_2^2-R_1^2\right)$ of the annular condensate with a
vortex array remains constant for all $\Omega\ge \Omega_{h}$. Since
the areal vortex density is $\Omega/\pi$,  the number of vortices in
the annular region itself is ${\cal N}_a = \Omega\left(R_2^2-R_1^2\right) =
\Omega\left( \Omega_h^2-1\right)/\lambda$, which increases linearly with
$\Omega$.  In contrast, the number of phantom vortices associated
with the irrotational flow is ${\cal N}_{\rm irr} = \Omega R_1^2$ so
that the effective total number of vortices becomes

\begin{equation}\label{nvortex}
{\cal N } _v={\cal N}_a + {\cal
N}_{\rm irr}  = \Omega R_2^2 = \Omega \left(\frac{\Omega^2 -1+
\Omega_h^2 -1}{2\lambda}\right).
\end{equation}

It is not
difficult to manipulate the expressions for $R_2^2\pm R_1^2$ to
obtain the mean radius $R \equiv  \frac{1}{2}\left(R_2 +
R_1\right)$ and the width $d\equiv R_2-R_1$ of the annular
region, for example
\begin{equation}\label{dsq}
d^2 = \frac{\left(\Omega_{h}^2 - 1\right)^2}{\lambda\left[ \Omega^2
-1 +\sqrt{\left(\Omega^2 - 1\right)^2- \left(\Omega_{h}^2 -
1\right)^2} \right]}.
\end{equation}
This expression reduces to $R_2^2$ in the limit
$\Omega\to \Omega_{h}$,  where the hole just forms, and $d$
decreases continuously as $\Omega$ increases, which is an
  obvious consequence
of the fixed area of the annular condensate.  For large $\Omega\gg
\Omega_{h}$, Eq.~(\ref{dsq}) shows that
\begin{equation}
   \label{d}
   d\approx
  \frac{\Omega_{h}^2-1}{\Omega\sqrt{2\lambda}}=\frac{\sqrt2}{\Omega}
  \left(\frac{3\sqrt\lambda \, g }{2\pi}\right)^{1/3},
\end{equation}
whereas Eq.~(\ref{sd}) shows that the mean radius becomes
\begin{equation}\label{R}
   R\approx \frac{\Omega}{\sqrt{2\lambda}},
\end{equation}
in the same limit.

The  energy $E'$ of the rotating condensate with a central hole
follows by integrating the corresponding chemical potential in
Eq.~(\ref{mulh}) using the
first thermodynamic relation in (\ref{thermo}) and noting that
$g\propto N$
\begin{equation}\label{elh}
E'= \frac{3N}{40\lambda} \,\left(\Omega_{h}^2-1\right)^2-
\frac{N}{8\lambda}
\left(\Omega^2-1\right)^2= \frac{3N}{10} \left(\frac{3\sqrt\lambda
\, g}{2\pi}\right)^{2/3} -
\frac{N}{8\lambda}
\left(\Omega^2-1\right)^2.
\end{equation}
    The corresponding angular momentum per particle follows from  the second
of Eqs.~(\ref{thermo})
\begin{equation}\label{llh}
\frac{L_{h}}{N} = \frac{\Omega\left(\Omega^2-1\right)}{2\lambda}.
\end{equation}
The first factor reflects the usual linear relation between the angular
velocity and the angular momentum.  The second factor
$(\Omega^2-1)/(2\lambda)$ is the mean squared radius from
Eq.~(\ref{sd}).  Comparison with Eq.~(\ref{nvortex}) for the
effective total number of vortices shows that the angular momentum 
per particle
per vortex is $0.5$ when the hole first appears, but it increases
monotonically toward $1.0$ for $\Omega \gg \Omega_h$.
This behavior is readily understood. For any annular TF
condensate with approximate solid-body rotation
$\bm v \approx \bm \Omega\times \bm r$, density $n(r)$
in the interval $0\le R_1\le r\le R_2$ and total number of
vortices ${\cal N}_v = \Omega R_2^2$,  the angular momentum
per particle per vortex  is  the average of $r^2/R_2^2$  with $n(r)$ 
as weight factor.

It follows
from Eq.~(\ref{dens1}) that the maximum  density occurs at the mean
squared radius
\begin{equation}
   r^2_{\rm max} =  \frac{1}{2}\left(
  R_2^2+R_1^2\right)= \frac{\Omega^2-1}{2\lambda},
\end{equation}
and has the
      dimensional value $n_{\rm max} = (\Omega_{h}^2-1)^2/(32\pi d_\perp^2
a\lambda)$.  The corresponding (dimensionless) healing length becomes
\begin{equation}\label{xi}
\xi = \left(\frac{1} {8\pi n_{\rm  max} a \,d_\perp^2}\right)^{1/2}  = \frac{2
\sqrt\lambda}{\Omega_{h}^2-1} = \left(\frac{2\pi}{3\sqrt\lambda
\,g}\right)^{1/3},
\end{equation}
    which is constant for $\Omega\ge \Omega_{h}$, namely
for the annular condensate with a dense vortex array.  In contrast,
Eq.~(\ref{d}) shows that the width $d$ of the condensate decreases
like $\Omega^{-1}$ for large $\Omega$.

  \subsection{Validity of the Thomas-Fermi approximation}

The Thomas-Fermi approximation assumes that the  healing length $\xi$
is much smaller than the width of the annulus $d$.  Since $\xi$ is
independent of  $\Omega$ in the present model, this condition will
fail for sufficiently large $\Omega$.  After some
algebra,  the condition $\xi^2 \ll
d^2 $  yields the   explicit restriction
\begin{equation}\label{tfconstr}
\Omega^2-1+\sqrt{\left(\Omega^2-1\right)^2 -
\left(\Omega_{h}^2-1\right)^2}\ll
\frac{\left(\Omega_{h}^2-1\right)^4}{4\lambda^2} = 4
\left(\frac{3\sqrt\lambda \,g}{2\pi}\right)^{4/3}.
\end{equation}
For large
$\Omega$,  the constraint (\ref{tfconstr})
    reduces to $\Omega^2 \ll 2[3\sqrt\lambda \,g/(2\pi)]^{4/3}$. If $\Omega^
2$  violates this restriction, the width of the condensate
becomes too small to
satisfy the TF approximation.   Since  $\xi$ also characterizes the
size of the vortex core,  the condition $\xi^2 \sim d^2$  means  that
a vortex no longer fits in the annular gap, suggesting that at large
angular velocities
the system will exhibit a transition to a configuration with pure irrotational
flow (a giant vortex).  This transition will be discussed in Sec.\ V.

The Thomas-Fermi approximation also
assumes that the healing length $\xi$  is small compared to the
circular-cell radius  $l=1/\sqrt\Omega$ in harmonic-oscillator units.
Equations (\ref{d}) and (\ref{xi}) together imply that

\begin{equation}
d\xi = \sqrt 2 l^2.
\end{equation}
    Thus $d/\xi = \sqrt2 l^2/\xi^2$, so that the
inequalities $\xi\ll d$ and $\xi\ll l$ are closely related.

\section{Numerical solutions}

In this section we  study rapidly rotating condensates numerically,
using the full Gross-Pitaevskii equation, and
compare our results directly to those derived analytically in the previous
section. The dimensionless  two-dimensional time-dependent
Gross-Pitaevskii equation is
\begin{equation}
  i \frac{\partial \psi}{\partial t} = \left[-\frac{1}{2} \nabla^2 +
  \frac{1}{2}(r^2+\lambda r^4) + g|\psi|^2
  - i \Omega \left(y\frac{\partial}{\partial x}-
  x\frac{\partial}{\partial y}\right) \right] \psi
\label{eq:GP}
\end{equation}
where   $r^2=x^2+y^2$, and $\psi$ is again
normalized to unity.

Equation (\ref{eq:GP}) can be solved numerically by propagating some initial
wavefunction in imaginary time, \ where one simply  makes the replacement
$t \rightarrow -it$. Then,
over an appropriate time scale, the system relaxes to the ground
state at the given angular velocity $\Omega$. For sufficiently large
angular velocities, this ground state will
contain one or more vortices,  ultimately  arranged in a triangular
lattice. In order to break any
residual symmetries in the system, we add random noise to the initial
wavefunction. In imaginary time, vortices  then  appear at the edge
of the condensate before penetrating into the bulk and relaxing into the
lattice, similar  to the real-time simulations of \cite{Kasa02}.
In some cases, more than one random initialization was tried in order to check
whether we have reached the true ground state~\cite{note1}.

\begin{figure}[here]
   \scalebox{.43}
   {\includegraphics{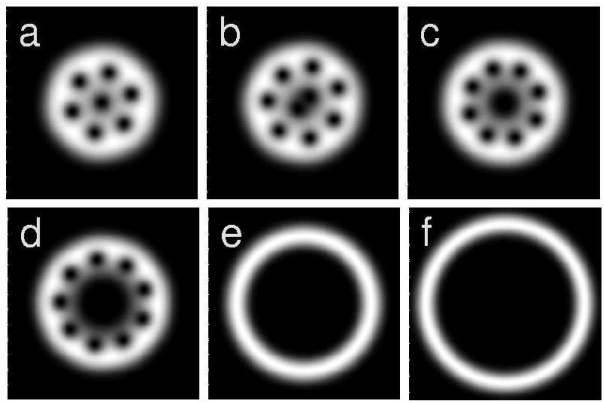}}
\caption{Density profiles of a rotating condensate at $g=80$ and
   $\lambda=0.5$, for (a) $\Omega=2.0$, (b) $\Omega=2.1$, (c) $\Omega=2.25$,
   (d) $\Omega=2.5$, (e) $\Omega=3.0$, (f) $\Omega=3.5$.  The scale of
   each figure is $4\times 4$ in units of $d_{\perp}$.}
\label{fig:vortglow}
\end{figure}

Figure \ref{fig:vortglow} shows the density $|\psi(x,y)|^2$ of the
condensate as
one increases the angular velocity, for $g=80$ and $\lambda=0.5$. For this
quite small interaction strength,  the
Thomas-Fermi approximation should not be particularly good (for
example, the nonzero density extends beyond the usual TF radii).  At
small
angular velocities [Fig.~\ref{fig:vortglow}(a)], one
  observes a vortex
lattice similar to that seen in
harmonic traps, with one singly quantized vortex at the center
surrounded by six others in a ring. As $\Omega$ increases, another
vortex appears near the central one [Fig.~\ref{fig:vortglow}(b)], until, for
$\Omega>2.25$ they merge to form a doubly quantized vortex surrounded by a
ring of singly quantized vortices [Fig.~\ref{fig:vortglow}(c)]. This
situation roughly corresponds to the case of
a vortex lattice with hole that is  expected in the
large-interaction limit. Then, as
$\Omega$ increases still further, the size of the hole and the
circulation around it both increase [Fig.~\ref{fig:vortglow}(d)].
Eventually,
the vortices in the outer ring recede into the hole.
Figure \ref{fig:vortglow}(e)  at $\Omega=3.0$ shows the resulting
density profile, where all
the vortices lie inside the hole. However, not all  of the
circulation is contained in a central multiply quantized vortex,
  since
other vortices are distributed around the center. Thus, this state cannot be
truly termed a giant vortex. Then as $\Omega$ increases further, the
circulation is entirely absorbed into the giant vortex, as shown in
Fig.\ \ref{fig:vortglow}(f) at $\Omega=3.5$. In this small-$g$
regime, the preceding discussion
highlights that the transitions between
the three phases (vortex lattice, lattice with hole, and giant vortex)
  are somewhat gradual. In Sec.\ V, we  use the Thomas-Fermi
approximation to   discuss the transition to the giant-vortex
state in more detail.

\begin{figure}[here]
   \scalebox{.43}
   {\includegraphics{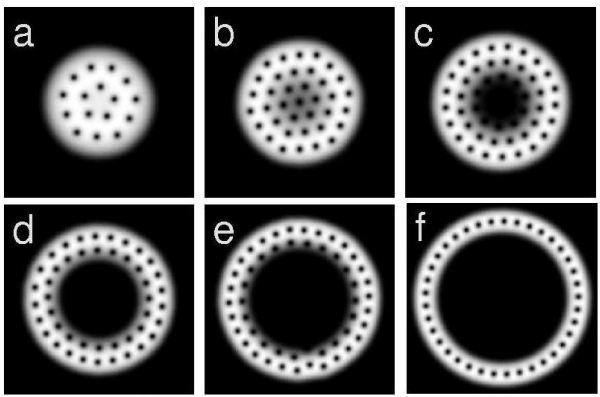}}
\caption{Density profiles of a rotating condensate at $g=1000$ and
   $\lambda=0.5$, for (a) $\Omega=2.0$, (b) $\Omega=3.0$, (c) $\Omega=3.5$,
   (d) $\Omega=4.0$, (e) $\Omega=4.5$, (f) $\Omega=5.0$, showing the stable
   vortice lattice configurations. The scale of each figure is $6\times
6$ in units of $d_{\perp}$.}
\label{fig:vortfm}
\end{figure}

We have also performed calculations for stronger interactions ($g=1000$).
In this case one finds a vortex lattice at $\Omega=2.0$, as
before~[Fig.~\ref{fig:vortfm}(a)].
However,  as  $\Omega$ increases, a density depression
appears in the center of the condensate, with an associated increase
in the core size of the
central vortices~[Fig.~\ref{fig:vortfm}(b)]. At approximately
$\Omega \simeq 3.3$, the density dips to
zero at the center, which can be taken to be the numerical value of
$\Omega_h$. Figure \ref{fig:vortfm}(c) shows a typical configuration with a
central hole surrounded by two rows of vortices. As $\Omega$  increases, the
inner row is absorbed by the expanding hole until, in Fig.\
\ref{fig:vortfm}(f)
($\Omega=5.0$) only one row of vortices remains. For larger
$\Omega$, we find that
the inner and outer radii of the condensate both increase in size,
but the basic
structure (a hole with a single ring of vortices) remains the same. We see no
transition to the giant-vortex
state up to $\Omega=7.0$, above which unfortunately the numerics become
very difficult. Nevertheless, this value sets a lower bound on
$\Omega_g$ for the transition to the giant vortex. For this angular
velocity, the TF gap width and healing length from Eqs.~(\ref{dsq})
and (\ref{xi}) are $d\approx 1.43$ and $\xi \approx 0.144$, so that
the TF approximation is well justified.  Also note that the density
distributions discussed here are
qualitatively similar to the radial profiles found in a
three-dimensional calculation
of Ref.\ \cite{Afta03}.

To make a quantitative comparison of the numerical
results in Fig.~\ref{fig:vortfm} with the analytical results, we have
measured  the intervortex spacing $b_{\rm meas}$ for the nearly
triangular lattices in Figs.~\ref{fig:vortfm}(a)-(d) and compared
them with the predicted values for an ideal triangular lattice using
the expression $b_{\rm tri}^2 = 2\pi/(\sqrt3\Omega)$.  Table I
shows that they agree very well, but  the measured values $b_{\rm
meas}$ decrease somewhat more slowly than would be expected for an
ideal lattice at the same angular velocity.

\begin{table}[here]
\caption{Comparison of  numerical values for intervortex lattice
spacing $b_{\rm meas}$ in a rotating condensate with $g=1000$ and
$\lambda = 0.5$  (obtained by direct measurements from
Fig.~\ref{fig:vortfm}) to the theoretical intervortex  spacing
$b_{\rm tri}$  for an ideal triangular lattice.}

\begin{ruledtabular}
\begin{tabular}{c|c|c}
\noalign{\vspace{.1cm}}
    % \hline   \noalign{\vspace{.1cm}}

$\Omega$&  $b_{\rm meas}$ &
$b_{\rm tri}$   \\
\noalign{\vspace{.1cm}}
    \hline \noalign{\vspace{.1cm}}
     2.0& 1.32 & 1.35 \\
       3.0 & 1.08 & 1.10 \\
     3.5 & 1.01& 1.02 \\
        4.0 & 0.97 &0.95 \\
\end{tabular}
\end{ruledtabular}
\end{table}

For larger angular velocities, the vortex array is  one-dimensional
and the annular condensate becomes increasingly narrow.
The
mean radius $R_{\rm meas}$  and width $d_{\rm meas}$ can be measured from
the numerical solutions by defining a reference density (in this case
$|\psi|^2=10^{-4}$) and noting the radii at which the density
is equal to this reference. Note that the inner and outer radii derived from
this
technique will differ from those calculated using the TF approximation, since
in the latter case the density vanishes at well-defined points, while for
the numerical results it does not. Nevertheless, the value of the measured mean
radius is not likely to be affected too much by this detail and
can be compared with the TF prediction
$R_{\rm TF}$, which follows from an expression similar to
Eq.~(\ref{dsq}).  In addition,  from $R_{\rm meas}$ and the number ${\cal
N}_a$ of vortices in the array, we can readily
  calculate the intervortex spacing
$b_{\rm meas}$  (Fig.\ \ref{fig:vortfm} contains only the condensate
for $\Omega = 5.0$, but those for $\Omega=6.0$ and 7.0 are
qualitatively similar).  Finally, the
measured width $d_{\rm meas}$ of the annular condensates can be compared to
the TF
prediction in Eq.~(\ref{dsq}). Note, however, that
$d_{\rm meas}$ is affected by the numerical technique used to extract the
radii, so only the $\Omega$ dependence can be checked when comparing to
$d_{\rm TF}$.  Table II contains these quantities.
The most striking feature is that the measured intervortex spacing
actually {\it increases} with increasing $\Omega$, in contrast to the
$1/\sqrt\Omega$  dependence for an ideal triangular lattice.  This
behavior, which can be considered an extrapolation of that in Table
I,  may be a precursor of the transition to a giant vortex. The values of 
${\cal N}_a$ found in our numerical
solutions are much
lower than those calculated from (\ref{difflh}), which may reflect a
non-uniform
vortex density near to the edge of the condensate. In addition, 
Eq.\ (\ref{difflh})
assumes a dense vortex array and therefore gives only a
semi-quantative estimate of the number of vortices
in the annulus when the array becomes one dimensional.

\begin{table}[here]
\caption{Comparison of selected  numerical values for a rotating
condensate with $g=1000$ and $\lambda = 0.5$ (obtained from
Fig.~\ref{fig:vortfm}) to some of the TF predictions.  The
intervortex spacing $b_{\rm meas}$ for the one-dimensional vortex
arrays follows from direct measurements of the mean radius $R_{\rm
meas}$ and the vortex number in the annulus ${\cal N}_a$. }

\begin{ruledtabular}
\begin{tabular}{c|c|c|c|c|c|c}
\noalign{\vspace{.1cm}}
    % \hline   \noalign{\vspace{.1cm}}

$\Omega$ &   $R_{\rm meas}$ & $R_{\rm TF}$ & ${\cal N}_a$ & $b_{\rm
meas}$ & $d_{\rm meas}$ & $d_{\rm TF}$
     \\
\noalign{\vspace{.1cm}}
    \hline \noalign{\vspace{.1cm}}

     5.0&  4.83 & 4.79 &37  & 0.821 & 2.32 & 2.06\\
       6.0 &  5.80 &
5.86 &  44 & 0.828 & 1.98 & 1.68 \\
     7.0 &    6.85 & 6.89 & 51 & 0.844 & 1.76 & 1.43 \\

\end{tabular}
\end{ruledtabular}
\end{table}

\begin{figure}[here]
   \scalebox{.43}
   {\includegraphics{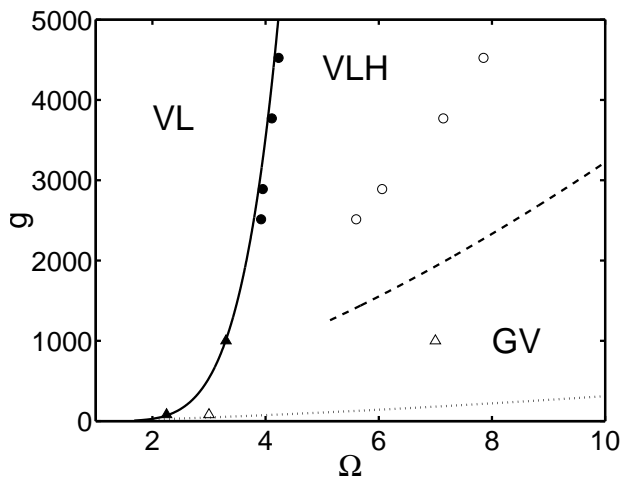}}
\caption{Phase diagram of a condensate rotating with angular velocity
   $\Omega$ versus interaction strength  $g$, with
  $\lambda=0.5$, where
   the solid line is the critical frequency $\Omega_h$
[Eq.~(\ref{omegalh})] derived from our TF analysis. Going
   from left to right, the first state (denoted VL) is a vortex
lattice without a hole,
   while above $\Omega_h$ (solid line) a hole appears (denoted VLH). Then, for
   $\Omega>\Omega_g$ (dashed line) the TF analysis predicts a giant
 vortex state (denoted GV). In addition, the dotted line indicates the
 crossover region where the TF approximation fails [defined here by
 the somewhat arbitrary  criterion $d^2 \approx  10\xi^2$, where $d$
 and $\xi$ are given by  Eqs.~(\ref{dsq}) and (\ref{xi})].  For
   comparision we also show the results of Kavoulakis and Baym
   \cite{Kavo03}, where the filled and open circles represent $\Omega_h$ and
   $\Omega_g$ respectively.  Finally, our numerical results for these 
 quantities are plotted as filled and open triangles.}
\label{fig:phased}
\end{figure}

Figure \ref{fig:phased} compares the $\Omega$-$g$ phase diagram resulting
from this numerical analysis to
the analytical TF results presented earlier. One can see that the value of
$\Omega_h$ found in the numerical calculations is very close to the
analytical result in Eq.~(\ref{omegalh}).  This is true not only for
the value of  $g=1000$
presented in Fig.~\ref{fig:vortfm};  even for $g=80$ the agreement
is very good,
which is somewhat surprising  since the Thomas-Fermi
approximation assumed in the analytic result is expected to be less accurate.
Given the difficulty in
identifying the exact point  at which the hole forms, the
numerical value of $\Omega_h$ quoted here is approximate, with an error of
$\pm 0.1$. In addition one can see that
both the analytical and numerical results are very close to those of
Kavoulakis and Baym \cite{Kavo03}.  Near the horizontal axis, Fig.\
\ref{fig:phased} also shows the crossover to a region where the
Thomas-Fermi approximation is expected to fail, as estimated by the
criterion $d^2\approx 10 \xi^2$ with $d$ and $\xi$  given by
Eqs.~(\ref{dsq}) and (\ref{xi}).

We have also calculated the energies of these vortex-lattice
configurations.  The numerical and analytical results will be
discussed  in the next section.

\section{Irrotational flow in annular region (``giant
vortex'')}\label{secirr}

   The  vortex array considered in Sec.~III
has the uniform vortex density $\Omega/\pi$ in dimensionless
oscillator units, which defines an effective vortex radius
$l=1/\sqrt\Omega$.  In any rotating superfluid, this quantity
decreases with increasing angular velocity.  The new feature of  the
present system is that the geometry itself also depends on $\Omega$.
In particular, Eq.~(\ref{d}) shows that the width of the annulus
decreases proportional to $1/\Omega$ and eventually becomes
comparable with the intervortex distance $\approx 2l$. Near this
critical angular velocity, the vortex array  first becomes one-dimensional
and then should disappear in a transition (perhaps a crossover) to
pure irrotational flow, as seen in the numerical
simulations of Sec.\ IV and Ref.~\cite{Kasa02} and studied in an
approximate
TF theory  in Ref.~\cite{Kavo03}.

\subsection{Pure irrotational flow}
\label{pureirr}

To study this behavior in detail, it is first
necessary to first consider the case of pure
irrotational flow, when
$\psi(\bm r) = e^{i\nu \phi}|\psi(r)|$ with $\nu$ the quantum of
circulation (and $\nu$ is  the angular momentum in units of
$\hbar$). The
resulting velocity is azimuthal with $v_{\rm irr}(r) = \nu/r$.  Since
$\nu$ will be large, it can be treated as a continuous variable,
ignoring the discrete transitions between adjacent large integral
values. The TF density is now given by

\begin{equation}\label{TFirrot}
g |\psi|^2 = \tilde \mu -U(r^2),
\end{equation}
   where $\tilde \mu = \mu + \Omega\nu $ is a
constant~\cite{Fisc03,Kavo03} and
   \begin{equation}\label{U}
U(x) = \frac{1}{2} \left(  \frac{\nu^2}{x} + x + \lambda x^2\right)
\end{equation}
   can be considered an effective potential that
combines the centrifugal barrier and the original trap potential
(here, $x =  r^2$).  This function  $U(x) $ has a single minimum at
$x_0$ determined by $U'(x_0)= 0$, so that  $x_0$ is also the
position of the local maximum in the TF density.  The density
vanishes at the  two classical turning points $x_1= R_1^2 $ and
$x_2=R_2^2$, determined by the condition

\begin{equation}\label{edge}
   U(x_i) = \tilde \mu ,
   \end{equation}
where $x_1 < x_0 < x_2$.

The normalization condition can be written
   \begin{equation}\label{normirr}
g =\pi \int_{x_1}^{x_2} dx\, \left[\tilde \mu  - U(x)\right],
\end{equation}
   along with the free energy per particle
   \begin{equation}\label{Firr}
\frac{gF}{N} = -\frac{1}{2}\pi \int_{x_1}^{x_2} dx\,\left[ \tilde \mu
-U(x)\right]^2.
\end{equation}
Here, the physical quantities
$N(\mu,\nu,\Omega)$ and $F(\mu,\nu,\Omega)$ depend explicitly on the
chemical potential $\mu$ and the angular velocity $\Omega$;
equivalently,  the normalization condition (\ref{normirr}) can also
be obtained from the thermodynamic derivative $N = -\partial
F/\partial \mu$.  The equilibrium value of the circulation $\nu$,
which is an additional parameter in the calculation, follows by
minimizing the free energy,
$\partial F/\partial \nu = 0$,
yielding~\cite{Kavo03}
\begin{equation}\label{nu}
g\,\Omega = \nu\pi \int_{x_1}^{x_2} \frac{dx}{x}  \left[\tilde
\mu  - U(x)\right].
\end{equation}

For a given $N$ and $\Omega$, Eqs.~(\ref{edge}),
(\ref{normirr}) and (\ref{nu}) must be evaluated to find the energy
$E'(N,\Omega)$ in the rotating frame.  Comparison with the
corresponding quantity for the condensate containing a
one-dimensional vortex array  then determines the phase diagram.  In
Ref.~\cite{Kavo03}, this procedure was carried out numerically, which
does not emphasize the relevant parameters needed for a physical
interpretation.  Instead,  we recall that  the   annulus expands
radially and becomes increasingly narrow for large $\Omega$, as seen
in Eqs.~(\ref{d}) and (\ref{R}).  Thus an expansion in  the small
parameter $d/R\sim [\lambda^2 \,g/(4\pi)]^{1/3}\Omega^{-2}$ characterizes
the irrotational state in  the relevant  limit of rapid rotation.

The minimum $x_0$ of the potential $U(x)$ in Eq.~(\ref{U}) satisfies
the cubic equation
\begin{equation}\label{cubic}
2\lambda x_0^3
+x_0^2 = \nu^2,
\end{equation}
which makes a direct analysis quite intricate.  Instead, the large
value of  $x_0 \approx R^2$ relative to $x_2-x_1 \approx 2Rd$ makes
it natural to expand $U(x)$ around $x_0$, and it is necessary to
include corrections through order $(x-x_0)^4$.  To make this
procedure  precise, let $x - x_0 = x_0 \delta$, where $|\delta|\ll
1$, so that the inner and outer boundaries are given by $x_i =
x_0(1+\delta_i)$.  In addition, let
\begin{equation}\label{delta0}
\tilde\mu -U(x_0) = \frac{1}{2} (3\lambda x_0^2 + x_0)\delta_0^2,
\end{equation}
    which defines $\delta_0$ in terms of the other parameters.  It is
not difficult to determine $\delta_1$ and $\delta_2$ as  series in
powers of $\delta_0$, and it is sufficient to keep terms of order
$\delta_0^3$.  In this way, Eqs.~(\ref{normirr}) and (\ref{Firr})
reduce to integrals of the form $\int_{\delta_1}^{\delta_2}
d\delta\,\cdots$, where $\cdots$ denotes a  function that is readily
expanded in powers of $\delta$.

A lengthy analysis eventually leads
to  useful and physical expressions.  For example, the ratio of
Eqs.~(\ref{normirr}) and (\ref{nu}) yields   the somewhat implicit
expression for the quantum of circulation
\begin{equation}\label{nu1}
\nu \approx \Omega x_0 \left(1 + \frac{\delta_0^2}{30\lambda x_0}\right).
\end{equation}
As expected, the leading term here is simply the
quantum number associated with the angular momentum $\Omega x_0 =
\Omega R^2$ of a narrow annulus, and the term of order
$\delta_0^2$  arises from the small TF corrections.

The next step is to solve the cubic equation (\ref{cubic})
   for $x_0$ by expanding it for large values of
$(\nu^2/2\lambda)^{1/3}$. The resulting  power series for
$x_0$  can be
combined with (\ref{nu1}) to give the mean squared radius

\begin{equation}\label{x0}
x_0 \approx \frac{\Omega^2 -1}{2\lambda } + \frac{\delta_0^2}{15\lambda},
\end{equation}
and the associated circulation
\begin{equation}\label{nu2}
\nu \approx \frac{\Omega\,(\Omega^2-1)}{2\lambda} +
\frac{\Omega\,\delta_0^2}{10\lambda}.
\end{equation}
Note that the leading term for $x_0$  reproduces
(\ref{sd}) for the mean squared radius of an annulus with a uniform
vortex array and the leading term for $\nu$ reproduces the angular
momentum (\ref{llh}) for an annulus with a uniform vortex array; in
both cases,  the corrections involve the small parameter $\delta_0^2$.
    A combination of Eqs.~(\ref{delta0}) and (\ref{nu2}) then gives the
chemical potential
\begin{equation}\label{muirr}
\mu  \approx  -\frac{(\Omega^2-1)^2}{8\lambda} + \frac{3\Omega^4
\delta_0^2}{8\lambda}.
\end{equation}

The remaining step is to determine the small
parameter $\delta_0$ from Eqs.~(\ref{normirr}) and (\ref{x0}),
leading to
\begin{equation}\label{delta1}
\delta_0 \approx \frac{2\sqrt \lambda}{\Omega^2}
\,\left(\frac{\sqrt\lambda\, g}{2\pi}\right)^{1/3}.
\end{equation}
Since $\delta_0$ is small, this result requires
$[\lambda^2 g/(4\pi)]^{1/3} \ll \Omega^2$, which ensures that $d/R\ll 1$.
These relations readily provide the inner and outer squared
radii  $R_1^2 =
x_0 + x_0\delta_1$ and $R_2^2 = x_0 + x_0\delta_2$.  To leading
order, the mean squared radius is $x_0$ and the width is given by $d^2
\approx x_0 \delta_0^2\approx (2/\Omega^2)[\sqrt\lambda
\,g/(2\pi)]^{2/3}$.   In more detail,
\begin{eqnarray}\label{Rp}
R_2^2 + R_1^2 & \approx & \frac{\Omega^2 - 1}{\lambda} +
\frac{4}{3\Omega^2} \left(\frac{\sqrt\lambda
\,g}{2\pi}\right)^{2/3},\\
\label{Rm}
R_2^2 - R_1^2 & \approx &
\frac{2}{\sqrt\lambda}\left(\frac{\sqrt\lambda \,g}{2\pi}\right)^{1/3}.
\end{eqnarray}
Comparison with Eqs.~(\ref{sd}) and (\ref{difflh})
shows that the mean squared radius here slightly exceeds that for the
annulus with a uniform vortex array, whereas the width here is
smaller by a factor $3^{-1/3}$.

Substitution into Eq.~(\ref{muirr}) expresses the chemical
potential for the irrotational giant vortex  in terms of the
appropriate variables $N$ and $\Omega$
\begin{equation}\label{muirr1}
\mu_{\rm irr} \approx   -\frac{(\Omega^2-1)^2}{8\lambda} +
\frac{3}{2}\left(\frac{\sqrt\lambda \,g}{2\pi}\right)^{2/3}.
\end{equation}
Integration of the thermodynamic relation
(\ref{thermo}) then yields the energy in the rotating
frame
\begin{equation}\label{Eirr}
E_{\rm irr}' \approx   - \frac{N}{8\lambda} \left(\Omega^2-1\right)^2+
\frac{9N}{10}\left(\frac{\sqrt\lambda \,g}{2\pi}\right)^{2/3}.
\end{equation}
It is not surprising that   the energy (\ref{elh}) for
an annulus with  uniform and continuous vorticity $2\bm\Omega$ is
lower than (\ref{Eirr})  for pure irrotational flow.  Indeed,  a
fluid in solid-body rotation necessarily has a lower energy in the
rotating frame than other velocity distributions,  which
follows by minimizing the functional $E'[{\bm v}] = \int dV\rho
\left(\frac{1}{2}v^2 -\bm{\Omega\cdot   r\times v } \right)$ with
respect to $\bm {v}(\bm r)$.

\begin{figure}[here]
   \scalebox{.43}
   {\includegraphics{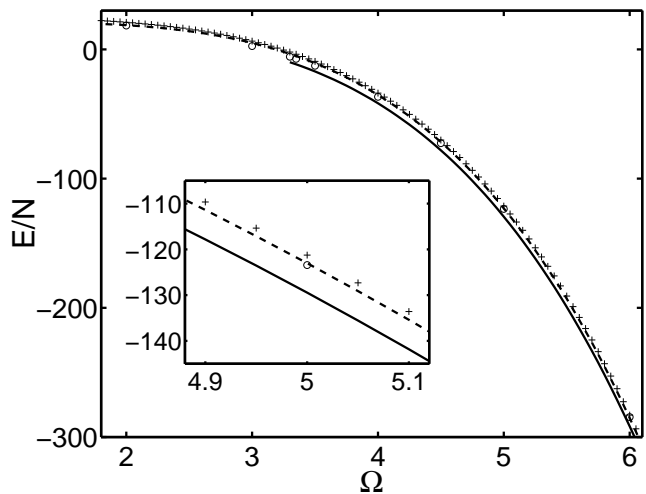}}
\caption{Energy per particle {\it vs.} rotation rate $\Omega$, where our
   numerical results are presented as open circles (vortex-lattice state) and
   crosses (giant-vortex state). For comparision, the analytical results are
   plotted as a solid line (Thomas-Fermi vortex lattice with hole) and a
   dashed line (giant vortex). The inset shows the energies in a small
   region of the graph around $\Omega=5$. The parameters used here are
   $g=1000$ and $\lambda=0.5$.}
\label{fig:energydiag}
\end{figure}

A good test of the analytical TF results is to compare the
TF energy per particle  to the values extracted directly from our
numerical calculations, as is shown in Fig.\ \ref{fig:energydiag}
for $g=1000$ and $\lambda = 0.5$. The open
circles are the energies corresponding to the vortex-lattice states
calculated numerically as shown in  Fig.\ \ref{fig:vortfm}. These
are compared to
the analytical result from Eq.~(\ref{elh}), which was
derived assuming a Thomas-Fermi profile and uniform vorticity. We find a
small but significant discrepancy between them, where the analytical result is
smaller by typically $\Delta E \sim 6$. This is not surprising
because (\ref{elh}) assumes solid-body rotation which  provides a lower
bound to the energy.  In detail, we attribute
this
discrepancy to the energy associated with the vortex cores, which will give a
correction to the analytical result due to the irrotational flow and density
dip near to each vortex.

\begin{figure}[here]
   \scalebox{.43}
   {\includegraphics{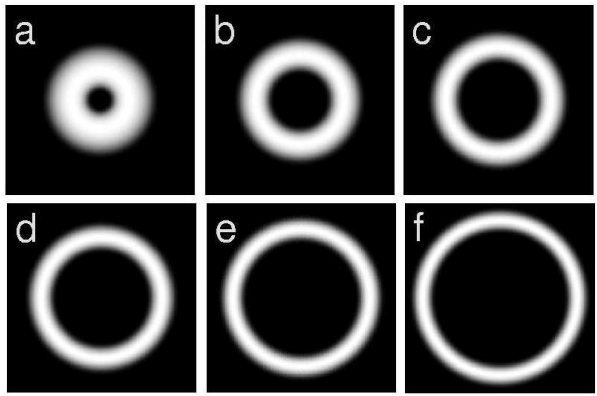}}
\caption{Density profiles of a rotating condensate at $g=1000$ and
   $\lambda=0.5$, showing the  giant-vortex states corresponding to the
   angular velocities  in Fig.\ \ref{fig:vortfm}:
     (a) $\Omega=2.0$, (b) $\Omega=3.0$, (c) $\Omega=3.5$,
   (d) $\Omega=4.0$, (e) $\Omega=4.5$, (f) $\Omega=5.0$. The scale of
each figure is $6\times 6$ in units of $d_\perp$.}
\label{fig:macrovortfm}
\end{figure}

We also calculate the energy of the
giant-vortex state by numerically solving Eq.\ (\ref{eq:GP}) with $\Omega=0$
and replacing $V_{\rm tr}$ with $V_{\rm tr}+\nu^2 /(2r^2)$, where $\nu$ is
the quantum of circulation  of the macroscopic flow. We
calculate the energy  $E_{\nu}$ for each
$\nu$, which in the rotating frame becomes
$E_{\nu}'=E_{\nu}-\nu \Omega$. The energy as a function of $\Omega$ is then
the minimum possible $E_{\nu}'$ for that particular $\Omega$, which also
fixes the relevant $\nu$. The density profiles of
the giant-vortex states corresponding to the same angular velocities as in
Fig.\ \ref{fig:vortfm} are plotted in Fig.\ \ref{fig:macrovortfm}, while the
energies are plotted
as the crosses in Fig.\ \ref{fig:energydiag}. Note that the energy of
the vortex lattice is always lower than the corresponding giant vortex
for the range of $\Omega$ covered, which reconfirms the conclusion
from Fig.\ \ref{fig:vortfm}
that no transition occurs to the giant-vortex state since it is energetically
unfavorable to do so.  It is also in contrast to the results at $g=80$,
which demonstrate that the giant vortex state can indeed become
energetically favorable
at high angular velocities (as seen in Fig.\ \ref{fig:vortglow}).
We also compare our results to the analytical energy
of the giant vortex Eq.\ (\ref{Eirr}) and find quite good agreement
for larger values of $\Omega$, with the analytical result being smaller by
around $\Delta E \sim 2$.

\subsection{Transition to a giant vortex}

The assumption of
uniform vorticity is reasonable as long as the intervortex distance
is much smaller than any of the dimensions of the annulus. It fails
when the discrete character of the quantized vortex lines becomes
significant, in particular, when the width $d$  of the annulus
becomes comparable to the intervortex separation $b\sim 2l =
2/\sqrt\Omega$.

  As background, it is instructive to review the analogous transition in
  rotating superfluid $^4$He  in an annulus. Here,
the width $d$ is
fixed by the geometry of the container, and vortices first appear
with increasing $\Omega$ when $b$ is of order $d$.
To be  more
precise, the irrotational flow in an annulus has the characteristic
feature that the velocity $v_{\rm irr} = \nu/r$ {\it decreases} with
increasing $r$, whereas the preferred solid-body flow $v_{\rm sb} =
\Omega r$ {\it increases}.  For a fixed gap width, this discrepancy
grows as $\Omega$ increases, and it eventually becomes favorable to
insert a one-dimensional array of quantized vortices at the midpoint
of the annulus.   On average, the circulating velocity around each
core combines with the overall irrotational flow to mimic more
closely $\bm v_{\rm sb}$, thus lowering the energy in the rotating
frame.
A detailed analysis of the free energy of the various
relevant states ~\cite{Fett67} predicted  that a one-dimensional
vortex array appears with {\it increasing} $\Omega$ at a
(dimensional) critical angular velocity $\Omega_0 = (\kappa/\pi
d^2)\ln(d/\xi)$, where  $d$ is the width of the annular gap and $\xi$
is the vortex core radius.  The theory indicated that  the vortices
appear one  by one in a very narrow interval of $\Omega$, rapidly
building up a one-dimensional vortex array in the center of the gap.
Experiments on superfluid $^4$He using second-sound attenuation
verified this predicted critical angular velocity  in considerable
detail~\cite{Bend67} (including the location at the midpoint of the
gap).

In the present case of a rotating annular condensate, the
width $d$ decreases like $1/\Omega$, which is faster than the
decrease of the intervortex spacing.  Hence the transition here is
expected to occur in the reverse order, with the irrotational giant
vortex  appearing for large $\Omega$.
The relevant comparison state
is  a one-dimensional  vortex array in addition to the irrotational
flow, as seen in Fig.\ \ref{fig:vortfm}(f).
For a preliminary
estimate,  it is convenient  to study only the case of a single
vortex located at the midpoint of the gap combined with macroscopic
circulation.  The analysis is lengthy and will not be given here
because  the resulting critical angular velocity does not agree well
with the numerical work.   Thus it suffices  merely to state that
the transition from one vortex with irrotational flow  to the pure
irrotational flow is predicted to  occur at a critical angular
velocity
   \begin{equation}\label{omega0}
\Omega_g \approx \frac{1}{\ln(d^2/\xi^2)} \left(\frac{\sqrt\lambda
\,g}{2\pi}\right)^{2/3},
\end{equation}
   as the angular velocity increases.

Figure \ref{fig:phased} also compares the critical angular velocity
$\Omega_g$ for the transition from the vortex
lattice
to the giant-vortex state. First, we see that our analytical
$\Omega_g$ is much larger than that of Ref.\ \cite{Kavo03} (open circles). In
addition, our numerical result (open triangle)  is much larger than
even our analytical
prediction (in Fig.\ \ref{fig:phased} we plot the lower bound on
$\Omega_g$ in the case of  $g=1000$ since the numerical methods place
constraints on using higher rotations, as discussed above).   Note that we plot
(\ref{omega0}) only for large $g$, since logarithmic accuracy is
expected to be poor for small $g$. Even for $\Omega$ larger
than $\Omega_g$, the lowest energy solution of the GP equation does not
correspond to a giant vortex,  which suggests
that for these values of $\Omega$
the giant vortex is stable against the formation of a single vortex, but is
unstable against the formation of an array of vortices (see Fig.\
\ref{fig:vortfm}).

\section{Discussion}

This work has
considered a two-dimensional rapidly rotating Bose-Einstein
condensate in a radial trap with both quadratic and quartic confining
components, which allows the external rotation $\Omega$ to exceed the
harmonic trap frequency $\omega_\perp$.  Both an analytical
Thomas-Fermi description and  a full numerical study of the
Gross-Pitaevskii equation show the formation of a central hole at
essentially the same  critical angular velocity $\Omega_h$ (the
analytical value apparently remains quite accurate even for small
values of the coupling constant $g = 4\pi N a/Z$ where the
Thomas-Fermi approximation
is less valid).  For larger angular velocities, the numerical work
for $g=80$
indicates a transition to a pure irrotational state (a giant vortex),
but our approximate analytical model predicts too small a value for
this transition.

Experimental work at the \'Ecole Normale
Sup\'erieure, Paris~\cite{Bret04,Stoc04} created a quartic
confinement with  a blue-detuned Gaussian  laser
  directed along the axial direction.  For their
trap, the nonrotating condensate was cigar shaped, and the strength
of the quartic admixture was $\lambda\approx 10^{-3}$.  For
dimensionless rotation speed $\Omega \sim 1$, the condensate was
nearly spherical, and it remained  stable  for    $\Omega\lesssim
1.05$.  Near the upper limit, the condensate exhibited a definite
local minimum in the central density, confirming the general features
of our TF analysis.  Throughout the stable range of $\Omega$, the
measured shape of the condensate was fit to the TF prediction, which
served as a direct determination of $\Omega$.  They also used
surface-wave spectroscopy~\cite{Zamb98,Cozz03} to provide an
independent measure of the angular velocity (even though the visible
number of vortices appeared to be too small for $\Omega\ge 1$).

These experiments  are very puzzling when compared to the previous
numerical studies~\cite{Kasa02,Afta03}, to previous analytical
studies~\cite{Kavo03} and to the present work. Why do the experiments
fail to reach higher angular velocities, when the simulations readily
reach the regime when the condensate becomes annular?  One possible
source of the discrepancy is that the experimental condensate is
definitely three-dimensional, whereas the simulations are
two-dimensional (apart from~\cite{Afta03}).  It is notable that even
the experimental papers~\cite{Bret04,Stoc04} suggest repeating the
experiments with a condensate that is tightly confined in the $z$
direction.  In addition, the low temperature of the experiments
  eliminates most dissipative processes that can equilibrate the system.
  This feature may make it difficult for the condensate to acquire more
  angular momentum as $\Omega $ increases.   In contrast, the
  numerical simulations work in
imaginary time for a given $\Omega$, which leads to a state that is
at least  a local minimum in the energy.

Another different question
concerns the transition to the giant vortex.  Our numerical
simulations indicate that this transition occurs well beyond our
analytical estimate $\Omega_g$ in Eq.~(\ref{omega0}).  This estimate
is obtained by comparing the energy of the irrotational giant vortex
with a similar irrotational state containing one additional quantized
vortex at the midpoint of the annulus.  Based on the numerical
simulations [especially Fig.\ \ref{fig:vortfm}(f)], it seems probable
that adding more quantized vortices in a one-dimensional ring and/or
allowing the radius of the ring to vary would
lower the energy of
this state relative to the irrotational state, because the
specific choice of one vortex at the midpoint seems somewhat
arbitrary for a dilute trapped condensate.  For superfluid $^4$He in
a rotating annulus,  the image vortices raise the energy if a vortex
approaches either wall, but such an effect is absent in the present
case.   If so, such an improved
analysis would increase the   estimated critical angular velocity
for the transition to a giant vortex.  It is also conceivable that
the mean radius of the one-dimensional vortex array simply shrinks
inside the inner TF radius $R_1$.  This latter situation would suggest
a crossover instead of a sharp transition, because the actual
condensate necessarily extends into the classically forbidden region.
These questions remain for future
investigation.

\acknowledgments

We are grateful to M.~Cozzini for
helpful comments.  This work originated in discussions at the Aspen
Center for Physics, and the Kavli Institute for Theoretical Physics (KITP)
provided an opportunity for many discussions and to prepare  the
manuscript. The work  at KITP was supported in part by the
National Science Foundation under
Grant No. PHY99-0794.  ALF is grateful
to the Laboratoire Kastler-Brossel, \'Ecole Normale Sup\'erieure, Paris
and to BEC-INFM, University of Trento for hospitality during
extended visits.


\begin{thebibliography}{99}
\bibitem{note1} To propagate Eq.\ (\ref{eq:GP}), we use a Fast Fourier
   Transform method, where the partial derivatives (in both the kinetic
   and angular momentum terms) are treated by Fourier transforming into
   momentum space. This method involves discretizing the wavefunction using
   a grid, where we typically use a grid spacing of
   $\Delta x = \Delta y = 0.08$.
\end{thebibliography}
\end{document}